\documentclass[12pt,a4paper]{article}
\usepackage{graphicx}

\begin{document}
\begin{center}
{\Large\bf $\kappa$-deformed Statistics and the Formation of}
\vspace{0.5cm}
 {\Large\bf a Quark-Gluon Plasma}

\vspace{1cm}

A.M. Teweldeberhan$^1$, H.G. Miller$^1$, and R. Tegen$^{1,\:2}$
\vspace{0.5cm}

{\itshape
$^1$Department of Physics, University of Pretoria, Pretoria 0002, South Africa
\vspace{0.5cm}

$^2$Department of Physics, University of the Witwatersrand,  P.O. WITS 2050, 
Johannesburg, South Africa}
\vspace{0.5cm}

\date{today}

\end{center}
\begin{abstract}
The effect of the non-extensive form of statistical mechanics proposed by 
Tsallis on the formation of a quark-gluon plasma (QGP) has been recently 
investigated in ref. \cite{1}. The results show that for small deviations 
($\approx 10\%$) from Boltzmann-Gibbs (BG) statistics in the QGP phase, the 
critical temperature for the formation of a QGP does not change substantially 
for a large variation of the chemical potential. In the present effort we use 
the extensive $\kappa$-deformed statistical mechanics constructed by Kaniadakis 
to represent the constituents of the QGP and compare the results with ref. [1].
\end{abstract}

\newpage

Substantial theoretical research has been carried out to study the phase 
transition between hadronic matter and the QGP. When calculating the QGP 
signatures in relativistic nuclear collisions, the distribution functions of quarks and gluons are traditionally described by BG statistics. In the past few years the non-extensive form of statistical mechanics proposed by Tsallis [2] has found applications in astrophysical self-gravitating systems [3], solar neutrinos [4], high energy nuclear collisions [5], cosmic microwave back ground radiation [6], high temperature superconductivity [7, 8] and many others. In these cases a small deviation of the Tsallis parameter, q, from 1 (BG statistics) reduces the discrepancies between experimental data and theoretical models. Recently Hagedorn's [9] statistical theory of the momentum spectra produced in heavy ion collisions has been generalized using Tsallis statistics to provide a good description  of {\it $e^+e^-$} annihilation experiments [10, 11]. Furthermore, Walton and Rafelski [12] studied a Fokker-Planck equation describing charmed quarks in a thermal quark-gluon plasma and showed that Tsallis statistics were relevant. These results suggest that perhaps BG statistics may not be adequate in the quark-gluon phase.

It has been demonstrated [13, 14] that the non-extensive statistics can be 
considered as the natural generalization of the extensive BG statistics in the presence of {\itshape long-range} interactions, {\itshape long-range} 
microscopic memory, or {\itshape fractal space-time} constraints.  It was 
suggested in [5] that the extreme conditions of high density and temperature in ultra relativistic heavy ion collisions can lead to memory effects and {\itshape long-range} color interactions. For this reason, the effect of the non-extensive form of statistical mechanics proposed by Tsallis on the 
formation of a QGP has been recently investigated in [1]. The results show that for small deviations ($\approx 10\%$) from BG statistics in the QGP phase, the critical temperature, $T_c$, for the formation of a QGP does not change substantially for a large variation of the chemical potential, $\mu$. This suggests that the critical temperature is to a large extent independent of the total number of baryons participating in the heavy ion collision responsible for the formation of the QGP.

Starting from the one parameter deformation of the exponential function 
$\exp_{\{\kappa\}}(x)=(\sqrt{1+\kappa^2 x^2}+\kappa x)^{\frac{1}{\kappa}}$, a 
generalized statistical mechanics has been recently constructed by Kaniadakis 
[15], which reduces to the ordinary BG statistical mechanics as the deformation 
parameter, $\kappa$, approaches to zero.  The difference between Tsallis and 
Kaniadakis statistics is the following: Tsallis statistics is non-extensive and reduces to BG statistics (extensive) as the Tsallis parameter, $q$, tends to one. On the other hand, Kaniadakis statistics is extensive and tends to BG statistics as the deformation parameter, $\kappa$, tends to zero.  In the present effort we use the extensive $\kappa$-deformed statistical mechanics constructed by Kaniadakis to represent the constituents of the QGP and compare the results with [1].

For a particle system in the velocity space, the entropic density in $\kappa$-deformed statistics is given by [15]
\begin{equation}
\sigma_{\kappa}(\bar{n})=-\int d\bar{n}\:\ln_{\{\kappa\}} (\alpha \, \bar{n})\:,
\end{equation}
where $\alpha$ is a real positive constant and $\kappa$ is the deformation 
parameter ($-1<\kappa<1$). As $\kappa\rightarrow0$, the above entropic density 
reduces to the standard Boltzmann-Gibbs-Shannon (BGS) entropic density if 
$\alpha$ is set  to be one. The entropy of the system, which is given by 
$S_{\kappa}=\int_{\Re} d^nv\,\sigma_{\kappa}(\bar{n})$, assumes the form

\begin{equation}
S_{\kappa}=-\frac{1}{2\:\kappa}\int_{\Re} d^nv\left(\frac{\alpha^\kappa} 
{1+\kappa}\,\bar{n}^{1+\kappa}-\frac{\alpha^{-\kappa}} 
{1-\kappa}\,\bar{n}^{1-\kappa}\right)
\end{equation}
and reduces to the standard BGS entropy $S_0=-\int_{\Re} d^nv \:[\ln(\alpha \, 
\bar{n})-1] \: \bar{n}$ as the deformation parameter approaches to zero. This 
$\kappa$-entropy is linked to the Tsallis entropy $S_q^{(T)}$ through the 
following relationship [15]:
\begin{equation}
S_{\kappa}=\frac {1} {2} \frac{\alpha^\kappa}{1+\kappa}\, 
S_{1+\kappa}^{(T)}+\frac{1}{2}\frac{\alpha^{-\kappa}}{1-\kappa}\, 
S_{1-\kappa}^{(T)}+const.
\end{equation}

For $\alpha$=1, the stationary statistical distribution corresponding to the 
entropy $S_{\kappa}$ can be obtained by maximizing the functional
\begin{equation}
\delta\:[S_{\kappa}+\int_{\Re} 
d^nv\:(\beta\,\mu\,\bar{n}-\beta\,\epsilon\,\bar{n})]=0.
\end{equation}
In doing so, one obtains
\begin{equation}
\bar{n}=\exp_{\{\kappa\}} \beta(\mu-\epsilon)\:,
\end{equation}
which reduces to the standard classical distribution as $\kappa\rightarrow 0$.

The entropic density for quantum statistics is given by [15]
\begin{equation}
\sigma_{\kappa} (\bar{n})=-\int d\bar{n} \: 
\ln_{\{\kappa\}}\left(\frac{\bar{n}} {1+\eta\,\bar{n}}\right)\,,
\end{equation}
where $\eta$ is a real number. After maximization of the constrained entropy or, equivalently, after obtaining the stationary solution of the proper 
evolution equation associated to (6), one arrives to the following distribution [15]:
\begin{equation}
\bar{n}=\frac {1} {\exp_{\{\kappa\}}\beta(\epsilon-\mu)-\eta}\:,
\end{equation}
where $\eta=1$ for $\kappa$-deformed Bose-Einstein distribution and $\eta=-1$ 
for $\kappa$-deformed Fermi-Dirac distribution.

If we use $\kappa$-deformed statistics to describe the entropic measure of the 
whole system, the distribution function can not , in general, be reduced to a 
finite, closed, analytical expression. For this reason, we use the 
$\kappa$-deformed statistics to describe the entropies of the individual 
particles, rather than of the system as a whole. The single particle 
distribution functions of quarks, antiquarks and gluons are given by  
\begin{equation}
\bar{n}_{Q(\bar{Q})}=\left\{\sqrt{1+\kappa^2\,T^{-2}\,(k\mp\mu_{Q})^2}+\kappa 
\, T^{-1}\,(k\mp\mu_{Q})+1\right\}^{-1} \end{equation}
and
\begin{equation}
\bar{n}_G=\left\{\sqrt{1+\kappa^2\,T^{-2}\,k^2}+\kappa \, 
T^{-1}\,k-1\right\}^{-1} \end{equation}
respectively.  In the limit $\kappa\rightarrow 0$ one recovers the corresponding BG quantum distributions for quarks, antiquarks and gluons (see (15)  and 
(16) in [1]).

The expression for the pressure is given by
\[
P_{QGP}=\frac {d_Q \, T} {2 \, \pi^2}\int_0^\infty  dk \, k^2 
\left(\frac{f_Q^\kappa-f_Q^{-\kappa}}{2\,\kappa}+\frac{f_{\bar{Q}}^\kappa-f_{\bar{Q}}^{-\kappa}}{2\,\kappa}\right)-
\]
\begin{equation}
\frac {d_G \, T} {2 \, \pi^2} \int_0^\infty dk \, k^2 \, 
\left(\frac{f_G^\kappa-f_G^{-\kappa}} {2\,\kappa}\right)-B\,,
\end{equation}
where
\begin{equation}
f_{Q(\bar{Q})}=1+\left[\sqrt{1+\kappa^2 \, T^{-2}(k\mp\mu_{Q})^2}+\kappa \, 
T^{-1}(k\mp\mu_{Q})\right]^{-\frac{1}{\kappa}}\,,
\end{equation}
\begin{equation}
f_G=1-\left[\sqrt{1+\kappa^2 \, T^{-2}\, k^2}+\kappa \, T^{-1} \, 
k\right]^{-\frac{1}{\kappa}}
\end{equation}
and B is the bag constant which is taken to be (210 MeV)$^4$. Equation (10)  
reduces to (13) in [1] as  $\kappa\rightarrow 0$.

The hadron phase is taken to contain only interacting nucleons and antinucleons 
and an ideal gas of massless pions. Since hadron-hadron interactions are of 
{\itshape short-range}, the BG statistics is successful in describing particle 
production ratios seen in relativistic heavy ion collisions below the phase 
transition. The interactions between nucleons is treated by means of a mean 
field approximation as in [1].

Assuming a first order phase transition between hadronic matter and QGP, one 
matches an equation of state (EOS)  for the hadronic system and the QGP via 
Gibbs conditions for phase equilibrium:
\begin{equation}
P_H=P_{QGP},\hspace{1cm} T_H=T_{QGP}, \hspace{1cm} \mu_H=\mu_{QGP}
\end{equation}
With these conditions the pertinent regions of temperature, $T$, and baryon 
chemical potential, $\mu$,  are shown in figure (1) for $\kappa$= 0, 0.23 and 
0.29. For $\kappa$=0.23 (see fig. 2), we obtain essentially the same phase 
diagram as in the case of Tsallis statistics with $q$=1.1.  Since both Tsallis 
and $\kappa$-deformed statistics are fractal in nature, we observe a similar 
flattening of the $T(\mu$) curves. This can be interpreted as follows: the formation 
of a QGP occurs at a critical temperature which is almost independent of the 
total number of baryons participating in heavy ion collision. The resulting insensitivity of the critical temperature to the total number of baryons presents a clear experimental signature for the existence of fractal statistics for the constituents of the QGP.

\newpage
\begin{figure}[ht] \centering
\begin{center}
\includegraphics [scale=0.5, angle=0]{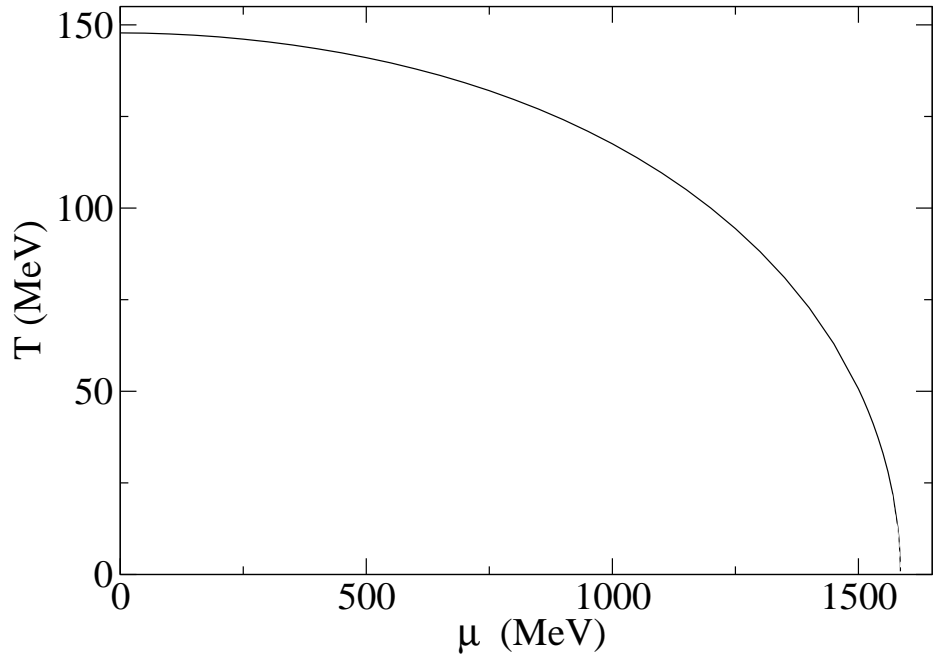}
\end{center}
\caption{Phase transition curves between the hadronic matter and QGP for 
$\kappa$=0 (solid line), $\kappa$=0.23 (dotted line) and $\kappa$=0.29 (dashed 
line).}
\end{figure}

\begin{figure}[ht] \centering
\begin{center}
\includegraphics [scale=0.5, angle=0]{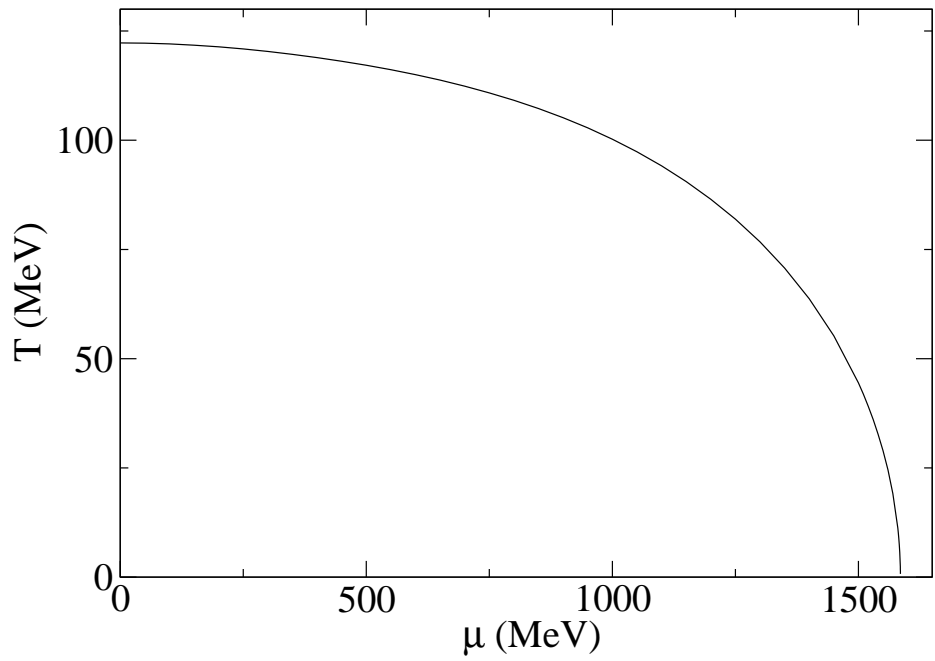}
\end{center}
\caption{Phase transition curves between the hadronic matter and QGP for 
$\kappa$=0.23 (dashed line) and $q$=1.1 (solid line).}
\end{figure}

\end{document}